\documentclass[12pt]{iopart}
\usepackage{iopams}
\newcommand{\eqref}[1]{(\ref{#1})}
\usepackage{amssymb}
\usepackage{hyperref}
\usepackage{tabularx}
\usepackage[dvipsnames]{xcolor}
\usepackage{soul}
\usepackage{siunitx} %
\usepackage{selinput}%
\usepackage{todonotes}
\usepackage[capitalise]{cleveref} %
\usepackage{upgreek} %

\usepackage{graphicx}
\usepackage{blindtext}
\def\newblock{\ } %
\usepackage{tabulary}
\usepackage{rotating}
\usepackage{ragged2e} %
\usepackage{afterpage}

\usepackage{multirow}
\usepackage{makecell}

\renewcommand{\thesection}{\arabic{section}}
\renewcommand{\thesubsection}{\thesection.\arabic{subsection}}
\renewcommand{\thesubsubsection}{\thesubsection.\arabic{subsubsection}}
\makeatletter
\newcommand*{\biggg}[1]{{\hbox{$\left#1\vbox to20.5\p@{}\right.\n@space$}}}
\newcommand*{\Biggg}[1]{{\hbox{$\left#1\vbox to23.5\p@{}\right.\n@space$}}}
\newcommand*{\Bigggg}[1]{{\hbox{$\left#1\vbox to26.5\p@{}\right.\n@space$}}}
\renewcommand{\p@subsection}{}
\renewcommand{\p@subsubsection}{}
\def\p@paragraph{\thesubsection.}
\makeatother

\Crefname{section}{Sec.}{Secs.}
\Crefname{table}{Tab.}{Tabs.}

\definecolor{mred}{RGB}{190,43,43}
\usepackage[normalem]{ulem}

\begin{document}
\title[Geometric tilt-to-length coupling in precision interferometry]{Geometric tilt-to-length coupling in precision interferometry: mechanisms and analytical descriptions}

\author{Marie-Sophie Hartig, S\"onke Schuster, and Gudrun Wanner}

\address{Max Planck Institute for Gravitational Physics (Albert Einstein Institute) and 
Institute for Gravitational Physics of the Leibniz Universit\"at Hannover, Callinstrasse 38, 30165 Hannover, Germany}

\ead{\mailto{marie-sophie.hartig@aei.mpg.de}, \mailto{gudrun.wanner@aei.mpg.de}}
\vspace{10pt}
\begin{indented}
\item[]January 2022
\end{indented}

\begin{abstract}
Tilt-to-length coupling is a technical term for the cross-coupling of angular or lateral jitter into an interferometric phase signal. It is an important noise source in precision interferometers and originates either from changes in the optical path lengths or from wavefront and clipping effects. Within this paper, we focus on geometric TTL coupling and categorise it into a number of different mechanisms for which we give analytic expressions. We then show that this geometric description is not always sufficient to predict the TTL coupling noise within an interferometer. We, therefore, discuss how understanding the geometric effects allows TTL noise reduction already by smart design choices. Additionally, they can be used to counteract the total measured TTL noise in a system. The presented content applies to a large variety of precision interferometers, including space gravitational wave detectors like LISA. 
\end{abstract}

\section{Introduction}
Precision laser interferometers often share a number of common noise sources, such as laser frequency noise, electronic readout noise, thermal noise, stray light and cross-talk.
There are numerous types of cross-talk since this term generally describes that a certain signal $s$ is picked up unintentionally by a sensor not built for sensing $s$. 
We extend this common definition slightly:
Let $e$ be an event describing changes in a degree of freedom (i.e.\ shifts or tilts). We speak of cross-talk or cross-coupling if $e$ leaks into a sensor readout that is not intended to detect $e$.

In the case of the interferometric phase, which should only read longitudinal distance variations between two reference points, any coupling of angular or lateral motion is therefore considered as cross-talk.
We refer to this cross-talk as tilt-to-length (TTL) coupling. 

The TTL coupling is a considerable noise source in high precision laser interferometry \cite{Troebs2018,pena2011mirror,hamann2019dilatometer,armano2016sub,Henry2020,sasso2018far-field,sasso2019}. 
In LISA Pathfinder, TTL coupling was visible as a `bump' in the noise spectrum at frequencies between \SIrange{20}{200}{mHz} %
\cite{mcnamara2008lisa,armano2016sub,armano2018cali,armano2019lpf} and was reduced by realignment and subtraction in data post-processing~\cite{Wanner2017}. In the second generation Gravity Recovery And Climate Experiment (GRACE Follow-On) \cite{sheard2012intersatellite,GFOabich19,Henry2020}, 
TTL coupling was considered as one of the highest noise contributors after laser frequency noise in the Laser Ranging Interferometer (LRI). In flight, the TTL noise was then shown to lie within the requirements \cite{Henry2020}.
Furthermore, TTL coupling is of particular interest in the future space gravitational wave detector LISA \cite{danzmann2011lisa,elisa13ARXIV,LISAMission,sasso2018misalignment,sasso2018far-field,sasso2019} where it is one of the most significant noise sources, and a variety of measures are being taken to suppress it optimally. Also, in other space gravitation wave detectors like Taiji and TianQin \cite{TianQinTaiji2021,Taiji-LISA2020,TianQin2016,Taiji2017}, TTL coupling will be a considerable noise source.

Within this paper, we systematically investigate a variety of TTL coupling mechanisms. We thereby focus on phase changes originating from alterations in the optical path length (OPL) of a laser beam, which can be described by trigonometric relationships. Therefore, we refer to these changes in the OPL as geometric TTL coupling.
These geometric TTL effects are categorised, described analytically and then classified as first- or second-order effects.

The total TTL coupling noise in a system is not described by OPL changes alone. We call any additional coupling contribution non-geometric TTL effects. These originate mostly from the properties of the interfering wavefronts as well as detector properties.
In this paper, we only demonstrate the relevance of the non-geometric effects and describe on one example how the full TTL signal (i.e.\ geometric + non-geometric) is computed. Additionally, we discuss here how the knowledge of the geometric TTL coupling can be used to mitigate also non-geometric TTL effects.
A detailed discussion of the non-geometric TTL effects is given in a follow-up paper \cite{NG21}.
The concepts discussed throughout this paper are fundamental and therefore independent of the application. They can be used in any laboratory experiment as well as in preparation for space missions such as LISA.

We introduce in \cref{Sec:TTL_in_different_systems} the different systems we consider and how angular and lateral jitter cause  OPL changes. In \cref{sec: all_geom} we categorise and model the various geometric TTL effects and reduce the equation for typical applications to first- or second-order effects. In \cref{sec: TTL vs geometry} we show exemplary systems, where a geometric TTL description is insufficient and fails to describe the interferometric phase readout. We summarise all effects in Sec.~\ref{Sec:GeomEffectsSummary+Mitigation}, list them for a typical special case and discuss how well-understood effects can be used to reduce the total TTL coupling within a system. Finally, we give a conclusion in Sec.~\ref{sec: summary}.

\section{TTL coupling in different systems due to angular and lateral jitter}
\label{Sec:TTL_in_different_systems}
\begin{figure}[tb]
	\centering
	\includegraphics[width= 0.7\columnwidth]{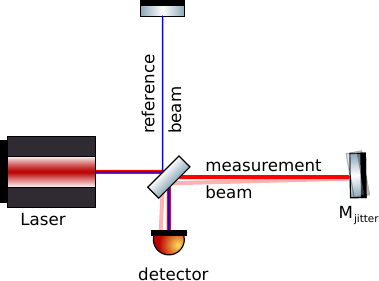}
	\caption{Michelson interferometer to illustrate angular cross-coupling -- the measurement beam is reflected from a tilting mirror M$_\text{jitter}$, while the reference beam does not change. This results in a misalignment of the two beams and a phase change of the measurement beam when impinging on the detector.}
	\label{fig: Michelson}
\end{figure}

Tilt-to-length coupling can occur in any type of interferometer. Only for initial illustration purposes, we assume a Michelson interferometer as depicted in Fig.~\ref{fig: Michelson}. Here, the incident laser beam is split into two arms, which we call a reference arm and a measurement arm. 
The reference beam, depicted in blue, is reflected from a hypothetically perfectly aligned plane mirror, transmits through the beam splitter and impinges orthogonally and centred on a photodetector. All components from which the reference beam reflects or through which it transmits are assumed to be static, which makes the reference beam equally static. 
The measurement beam is reflected first from a tiltable mirror M$_\text{jitter}$, and then from the beam splitter and impinges at an arbitrary angle and possibly off-centred on the detector.
\begin{figure}[tb]
	\centering
	\includegraphics[width= 0.45\columnwidth]{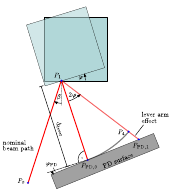} \hspace{3ex}
	\includegraphics[width= 0.45\columnwidth]{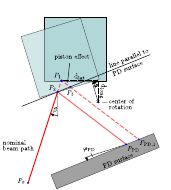}
	\caption{Geometrical cross-coupling splitting into a `lever arm effect'  and `piston effect'. 
Left figure - the lever arm effect: A beam impinges at a mirror at an angle $\beta$, is reflected in the point $P_1$, and then propagates the distance $d_\text{lever}$ until hitting the detector surface in the point $P_{\text{PD},0}$. The detector surface normal is tilted by an angle $\varphi_\text{PD}$ with respect to the nominal beam axis. 
	Tilting the mirror by an angle $\varphi$ around the beam's point of reflection $P_1$ yields a beam axis rotation of $2\varphi$ and causes the path length to increase by the distance from $P_4$ to $P_\text{PD,1}$. We call this path length change the lever arm effect.
	Right figure - the piston effect: If the mirror rotates around a centre that is shifted laterally by $d_\text{lat}$ and longitudinally by $d_\text{long}$ against $P_1$, diverts the beam from the dashed to the solid red axis. Thereby,
	 the beam's OPL reduces by the distance $P_2$ to $P_1$ to $P_3$ due to a shift of the mirror surface into (or out of) the beam path. This path length change is called the piston effect and is negative in this figure. The beam's point of incidence at the detector changes to $P_\text{PD}$. Arrows pointing clockwise indicate negative angles.}	
	\label{fig:geoTTL}
\end{figure}%
At the detector, both beams then interfere. Distance variations in the interferometer are then observed by reading the interferometric phase signal, which is often converted into units of lengths. We refer to this signal as the longitudinal path length sensing (LPS) signal. The LPS signal obtained without tilt (nominal case) 
\begin{equation}
	\text{LPS}_\text{nom} = \text{OPL}_{m,\text{nom}} - \text{OPL}_{r,\text{nom}} + \text{LPS}_\text{ng,\text{nom}}\;.
	\label{eq:LPSnom}
\end{equation}
and for a tilted mirror
\begin{equation}
	\text{LPS}_\text{tilt} = \text{OPL}_{m,\text{tilt}} - \text{OPL}_{r,\text{tilt}} + \text{LPS}_\text{ng,tilt}\;.
\end{equation}
contain each the difference of the OPLs of the measurement ($m$) and reference beam ($r$), so their optical path length difference (OPD). Additionally, the LPS signal contains a non-geometric contribution (LPS$_\text{ng}$) originating from the wavefront and detector properties.  
The reference beam is unaffected by the mirror jitter as visible in Fig.~\ref{fig: Michelson} such that its OPL does not change: $\text{OPL}_{r,\text{nom}} =\text{OPL}_{r,\text{tilt}}$. 
The TTL effect due to the mirror tilt is then found by comparing the tilted with the nominal case: 
\begin{eqnarray}
	\text{LPS}_\text{TTL} &= \text{LPS}_\text{tilt} - \text{LPS}_\text{nom} \\
	&= \text{OPL}_{m,\text{tilt}} - \text{OPL}_{m,\text{nom}} + \text{LPS}_\text{ng,tilt} - \text{LPS}_\text{ng,nom}\;. \\
	&= \text{OPD} + \text{LPS}_\text{ng,TTL} \label{eq:TTL_ng+g}
\end{eqnarray}
The geometric TTL effect is therefore entirely obtained from OPL changes of the measurement beam, and no information of the reference beam or its propagation path through the interferometer is needed. The OPD discussed within this paper thus always refers to the difference of the measurement beam's OPLs in a tilted (or shifted) and in the nominal case.
As stated above, we focus in this paper on the contribution of the OPL changes, while the non-geometric effects ($\text{LPS}_\text{ng,TTL}$) are discussed separately in \cite{NG21}.

We now revisit the setup depicted in Fig.~\ref{fig: Michelson} which is more complex than needed for TTL estimation. As we have seen, the reference path is of no relevance for TTL computations. Furthermore, the path of the measurement beam can be simplified.
Any reflection of the measurement beam from static flat components (i.e. planar mirrors or reflecting beam splitters) can be neglected since it only changes the beam path but not its path length. Hence, we unfold the reflected measurement beam in this reflection point. 
The detector is then placed into the unfolded beam path at the distance of its nominal path length from the tilted mirror. 
This reduces the setup to only the tiltable mirror and the photodiode, as shown in Fig.~\ref{fig:geoTTL}.

This reduction of the setup significantly simplifies TTL investigations. Moreover, it implies that the TTL estimation is independent of the interferometer type (Michelson, Mach-Zehnder, or others), because these are distinguished by the different routings of the measurement and reference beam.

The result, including a description of the alignment parameters and the occurring TTL effects, is depicted in Fig.~\ref{fig:geoTTL}. 
There, the beam nominally impinges on the mirror with angle $\beta$ at point $\text{P}_1$, is reflected and hits the photodiode at $\text{P}_{\text{PD,0}}$. Thereby, we assume $-90^\circ < \beta < 90^\circ$ to ensure that the beam impinges on the mirror's front surface.
If the mirror then rotates by an angle $\varphi$ around an arbitrary centre of rotation, the beam is reflected instead at point $\text{P}_2$ and hits the photodiode at $\text{P}_{\text{PD}}$.
The mechanisms leading to this beam path change are denoted as lever arm and piston effect. 
The lever arm effect describes the path length change due to the rotation of the reflected beam axis, while the piston effect accounts for the additional changes due to the rotation of the reflective surface into or out of the beam path. 
We will analyse both effects in more detail in the following section. 

The beam path after the rotation has a different path length than before, which means the OPL is angle-dependent. 
Therefore, we expect that any angular jitter of the mirror would result in phase noise. This type of effect is the most obvious TTL contribution: geometric TTL coupling due to angular jitter of a component resulting in OPL variations.

\begin{figure}[tb]
	\centering
	\includegraphics[width=1\columnwidth]{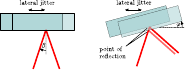}
	\caption{Sensing cross-talk due to lateral jitter of a mirror. Left: For a perfect mirror (no surface roughness or defects), the shown mirror motion does not affect the beam path, and there is no cross-talk. Right: The lateral jitter of the tilted mirror alters the beam path, resulting in a path length change that closely relates to the piston effect (see Fig.~\ref{fig:geoTTL}).}
	\label{fig: shift lateral}
\end{figure}

Also, lateral jitter can cause OPL changes, as depicted in Fig.~\ref{fig: shift lateral}. Since this effect occurs only if the mirror is tilted, we likewise call this effect a ``tilt-to-length'' coupling effect. However, in this case, the tilt is assumed to be static. 
This effect is related to the angular jitter case, particularly the piston effect, as we will show in Sec.~\ref{sec: geom} below.

Besides mirror jitter, there is another important jitter causing TTL coupling, which we refer to as receiving system jitter and discuss as a second application throughout this document. 
We speak of receiving system jitter (or receiver jitter) when the measurement beam propagates from one optical bench to another, and the entire receiving bench jitters. This is, for instance, the case when a beam is sent from one spacecraft to another, and the receiving spacecraft jitters relative to the arriving wavefront. It occurs, for example, in the gravity recovery mission GRACE Follow-On or in space gravitational wave detectors such as LISA, Taiji or TianQin.

Please note that the term `receiver' refers to the entire receiving system (e.g. spacecraft) in this work, not to a photoreceiver, which we call photodiodes here. 
We assume in this scenario that all components of the receiver system are rigidly connected to the jittering optical bench and therefore jitter with the optical bench but not with respect to each other. The reference beam is assumed to originate from a fibre injector located on the jittering bench or it is directly generated on the jittering bench. Thereby, the reference beam moves synchronously with all local components and impinges on the receiving photodiode always at the same angle and incidence point and with the same optical path length. We can, therefore, neglect the reference beam with all its properties from TTL computations (compare Eqs.~\eqref{eq:LPSnom}-\eqref{eq:TTL_ng+g}).

Like in the case of mirror jitter, we can neglect all reflective planar components and unfold the measurement beam's path. This is possible since these components divert the beam but do not affect the optical path length. 
Thereby, the receiver system reduces to a photodiode which is jittering synchronously with the receiving system and relative to the incident beam.

\begin{figure}[htb]
	\centering
	\includegraphics[width=0.7\columnwidth]{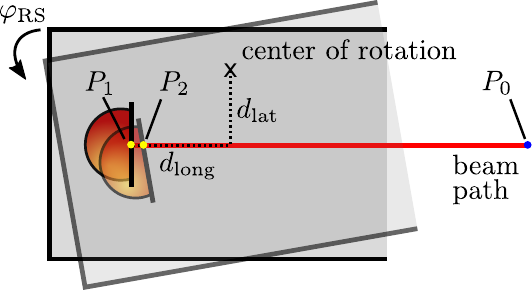}
	\caption{Geometrical cross-coupling due to angular jitter of the receiving system (grey open box) with respect to the incoming beam (red trace). In the untilted case, the beam originating from point $P_0$ hits the detector in point $P_1$. When the receiver rotates by the angle $\varphi_\text{RS}$, the beam hits the detector in point $P_2$ instead. 
	The distances $d_\text{long}$ and $d_\text{lat}$ define the longitudinal and lateral distances between the nominal point of incidence $P_1$ and the centre of rotation. Both are positive in this figure. 
	Switching their signs, i.e.\ placing the centre of rotation behind the photodiode ($d_\text{long}<0$) and below the beam axis ($d_\text{lat}<0$), switches also the sign of the path length changes. }
	\label{fig:geoSCTTL}
\end{figure}

This scenario is depicted in \cref{fig:geoSCTTL}. We see that a rotation of the system can move the photodiode position with respect to the beam and therefore change the beam's OPL. 
Like in the case of a reflection from a tilted mirror (see \cref{fig: shift lateral}), lateral receiver jitter can cause additional TTL coupling even if the angle indicated in \cref{fig:geoSCTTL} is static.
We will quantify and describe the different kinds of TTL effects below.

\section{Geometric TTL coupling effects}
\label{sec: all_geom}
Geometric TTL coupling describes the path length changes of the beams as depicted in \cref{fig:geoTTL} and \ref{fig:geoSCTTL}. 
We compare in this section the OPL of the nominal measurement beam with the optical path length the beam accumulates in the case of a jitter of the mirror or setup, respectively. %
We will derive the OPD for the case of a rotating mirror in \cref{sec: geom} and describe the other case of the rotating setup in \cref{sec: SCgeom}. 
Both will be compared in Sec.~\ref{sec: compare_SC_TM}.
Furthermore, we will extend our analysis to a broader set of cases in Sec.~\ref{sec:geom:extensions}, e.g., we will investigate OPDs due to transmissive components along the beam path, additional misalignments and three-dimensional cases.

\subsection{Geometric TTL coupling effects for a reflection at a mirror} 
\label{sec: geom}
We consider first the setup depicted in Fig~\ref{fig:geoTTL}. %
In the left part of this figure, the mirror rotates by an angle $\varphi$ around the reflection point $\text{P}_1$. 
As described in \cref{Sec:TTL_in_different_systems}, the beam then propagates an additional distance given by the geometrical distance between the points $\text{P}_4$ and $\text{P}_{\text{PD,1}}$. We call this geometric effect the lever arm effect. \\
If the mirror does not rotate around the reflection point but around an arbitrary centre of rotation the beam reflects instead at point $\text{P}_2$ and hits the photodiode at $\text{P}_{\text{PD}}$. 
We thereby assume that the centre of rotation is shifted longitudinally by $d_\text{long}$ and laterally by $d_\text{lat}$ against the nominal reflection point $P_1$.
In this case, the distance between the points $\text{P}_3$ and $\text{P}_{\text{PD,}1}$ is identical to the distance between $\text{P}_2$ and $\text{P}_{\text{PD}}$. Therefore, the only additional path length change is given by the distances between $\text{P}_2$, $\text{P}_1$ and $\text{P}_3$. We define this additional path length change due to an arbitrary centre of rotation as \textit{piston effect} since the reflecting surface moves in and out of the beam like a piston.

By assigning a position vector $\vec{p}_i$ to each point P$_i$, we can express the beams accumulated optical path length between its defined origin $P_0$ and the incidence point P$_{\text{PD,0}}$ on the photodiode as
\begin{equation}
	\text{OPL} (\varphi=0) = |\vec{p}_1 - \vec{p}_0| + |\vec{p}_\text{PD,0} - \vec{p}_1|\,,
	\label{eq:OPL(varphi)-zero}
\end{equation}
likewise the OPL in the rotated case can be expressed by 
\begin{equation}
	\text{OPL}(\varphi) = |\vec{p}_2 - \vec{p}_0| + |\vec{p}_\text{PD} - \vec{p}_2|\,.
	\label{eq:OPL(varphi)-general}
\end{equation}
While Eq.~\eqref{eq:OPL(varphi)-general} is naturally valid for all angles $\varphi$ including zero, we deliberately define here  Eq.~\eqref{eq:OPL(varphi)-zero} independently for a clearer description of the equations below.\\
We are now interested in the optical path length change due to the rotation, so we compute the optical path length difference (OPD):
\begin{equation}
	\text{OPD} = \text{OPL}(\varphi) - \text{OPL}(\varphi=0)\,. 
\end{equation}
From Fig.~\ref{fig:geoTTL} we can see that the OPD is the sum of the lever arm and piston effect and naturally independent of the choice of the starting point P$_0$:%
\begin{eqnarray}
	\text{OPD} &=	\left( |\vec{p}_\text{PD,1} - \vec{p}_4|\right) + 
				\left( |\vec{p}_2 - \vec{p}_1| + |\vec{p}_\text{3} - \vec{p}_1| \right) \\
			&=:	\text{OPD}_\text{lever} + \text{OPD}_\text{piston} \,.
				 \label{eq: piston fig} 
\end{eqnarray}
We use the fact that the distance between P$_1$ and P$_\text{PD,0}$ is identical to the distance between P$_1$ and P$_4$, and find
\begin{eqnarray}
	 \text{OPD}_\text{lever} &= 
	 |\vec{p}_\text{PD,1} - \vec{p}_1| - |\vec{p}_\text{PD,0} - \vec{p}_1|  \\
	 \text{OPD}_\text{lever}^\text{2D} &= 
	 d_\text{lever} \left[\sec(2 \varphi- \varphi_\text{PD} )\cos(\varphi_\text{PD}) - 1\right] \,. \label{eq: lever arm}
\end{eqnarray}
Eq.~\eqref{eq: lever arm} holds for two-dimensional investigations of this effect. Here and in the following, the index `2D' will denote the coupling in two-dimensional planes while the index `3D' refers to coupling in three dimensions.

For the piston effect, only absolute length changes are described by Eq.~\eqref{eq: piston fig}, however, the path length can increase or decrease in comparison to the non tilted case $\varphi =0$. If we account for this sign, we find:
\begin{eqnarray} 
	\text{OPD}_\text{piston}^\text{2D} =\, &2 \sec(2 \varphi-\varphi_\text{PD}) \cos(\beta+\varphi-\varphi_\text{PD})  \nonumber\\
& \cdot	\left\lbrace -d_\text{lat} \sin(\varphi) + d_\text{long}\left[1-\cos(\varphi)\right]\right\rbrace\,.
\label{eq: piston phi}
\end{eqnarray}
Here, $d_\text{long}$ and $d_\text{lat}$ are displacements of the centre of rotation from the nominal reflection point P$_1$ in longitudinal (i.e.\ here orthogonal to the reflecting surface) and lateral direction (here parallel to the reflecting surface). 
As before, $d_\text{long}$ is defined to be positive if the incoming beam passes the centre of rotation first and negative otherwise. Further, $d_\text{lat}$ is positive if the beam passes this centre to the left. Given the setup of Fig.~\ref{fig:geoTTL}, both distances are positive.
We assume that the interferometer is set to measure mirror displacements in the longitudinal direction, while motion in the lateral direction or angular jitter are by definition cross-talk -- and we will focus only on this cross-talk.

So far, we have defined variables without time dependency, but have implicitly assumed that $\varphi=\varphi(t)$ (i.e.\ the mirror performs angular jitter), while there is no lateral motion (i.e.\ $ d_\text{lat}=\text{const.}$). Likewise, we could assume that the mirror jitters in lateral direction such that $d_\text{lat} = d_\text{lat}(t)$.
In the latter case, the piston effect describes now lateral jitter coupling. Since the very same equation for the piston effect is being used and only interpreted differently, we do categorise lateral jitter coupling as a TTL effect. 
This is particularly visible in Fig.~\ref{fig: shift lateral} and also Eq.~\eqref{eq: piston phi}. If the mirror was ideally aligned ($\varphi = 0$), the lateral jitter would not couple. \\
Equations \eqref{eq: lever arm} and \eqref{eq: piston phi} are valid both for small and large angles $\varphi$. In our typical applications, we have a very small angular jitter, and the system is also well aligned. A Taylor series to second-order around zero is therefore useful for many applications:
\begin{eqnarray}
	\text{OPD}_\text{lever}^\text{2D} &\approx  ~2  d_\text{lever} \left[\varphi^2-\varphi\varphi_{\text{PD}} \right] \label{eq: lever sec order}\\
	\text{OPD}_\text{piston}^\text{2D} &\approx -2 d_\text{lat} \left[\cos(\beta)+\varphi_\text{PD}\sin(\beta)\right] \varphi  \nonumber\\
	&\; + \left[ 2 d_\text{lat} \sin(\beta)+d_\text{long} \cos(\beta) \right] \varphi^2 \,.
	\label{eq: piston sec order}
\end{eqnarray}
Especially for the case of normal incidence or nearly normal incidence ($\beta \approxeq  0, \varphi_\text{PD} \approxeq  0$), the equations reduce to
\begin{eqnarray}
	\text{OPD}_\text{lever}^\text{2D} &\approx ~2  d_\text{lever} \varphi^2 \label{eq: lever sec order simple}\\
	\text{OPD}_\text{piston}^\text{2D} &\approx -2 d_\text{lat}  \varphi  +d_\text{long} \varphi^2 \,.
	\label{eq: piston sec order simple}
\end{eqnarray}
In this special case, we find that the lever arm effect is purely second-order, while the piston effect splits into a first-order due to the lateral displacement and a second-order due to the longitudinal displacement of the component's centre of rotation with respect to the beam reflection point on the component. \\
This finding is of particular relevance for suppressing TTL coupling in typical optical setups:
Since these two geometric effects are known and well understood as described above, they can be used to counteract any unknown TTL effect, originating, for instance, from the non-geometric coupling. We will further discuss this in \cref{Sec:GeomEffectsSummary+Mitigation}.

\subsection{Geometric TTL coupling effects for rotating systems}
\label{sec: SCgeom}
Let us now assume a case of a freely propagating laser beam that is perfectly stable and does not jitter in any degree of freedom. This beam is then incident on a jittering receiving system, e.g. an optical table in a laboratory setup or a remote satellite in a space mission like GRACE-FO, LISA, Taiji or TianQin. 
We now assume the receiving system (i.e. optical bench or satellite) to jitter relative to the incident beam. In this case, the reference beam as well as all other possibly existent components such as mirrors, beamsplitters and photodiodes move perfectly synchronously, such that the OPL of the reference beam is constant. However, the photodiode then moves in and out of the received beam, as indicated in \cref{fig:geoSCTTL}, resulting in an OPL change, and therefore a geometric TTL coupling.

The geometric TTL coupling can easily be calculated by comparing the path lengths of the beam for the non-rotated and the rotated case. While the OPL for the nominal, non-rotated case is given by
\begin{eqnarray}
\text{OPL}_\text{RS}(\varphi_\text{RS}=0) = \vert \vec{p}_1 - \vec{p}_0 \vert \, ,
\end{eqnarray}
the OPL for the rotated case, i.e. $\varphi_\text{RS}\neq 0$, is in Fig.~\ref{fig:geoSCTTL} shorter, namely
\begin{eqnarray}
\text{OPL}_\text{RS}(\varphi_\text{RS}=0) = \vert \vec{p}_2 - \vec{p}_0 \vert \, .
\end{eqnarray}
The OPD is then given by the difference between the rotated and the non-rotated case, i.e.
\begin{eqnarray}
\text{OPD}_\text{RS} = \vert \vec{p}_2 - \vec{p}_0 \vert - \vert \vec{p}_1 - \vec{p}_0 \vert \, .
\end{eqnarray}
This OPD can be described analytically by the rotation angle $\varphi_\text{RS}$, and the distance between the point of detection and the centre of rotation. 
The absolute longitudinal difference, $d_\text{long}$, is in this paper defined by the distance between both points parallel to the nominal beam propagation axis. Meanwhile, the sign of $d_\text{long}$ is positive if the centre of rotation lies before the point of detection and negative otherwise. 
Analogously, the absolute lateral difference, $d_\text{lat}$, is given by the distance between both points along the axis orthogonal to the beam propagation axis. We define $d_\text{lat}$ to be positive if the beam ``sees" it at the right-hand side and negative otherwise, e.g. like in the case depicted in \cref{fig:geoSCTTL}.
With this, the OPD is given by
\begin{eqnarray}
\text{OPD}_\text{RS}^\text{2D} = \sec(\varphi_\text{RS}) \left\lbrace -d_\text{lat} \sin(\varphi_\text{RS}) + d_\text{long}\left[1-\cos(\varphi_\text{RS})\right]\right\rbrace\,,
\label{eq: SC OPD}
\end{eqnarray}
or, allowing the photodiode to be tilted by an angle $\varphi_\text{PD}$ like in the case of a rotating mirror (compare Fig.~\ref{fig:geoTTL}), we have
\begin{eqnarray}
\text{OPD}_\text{RS}^\text{2D} &=
 -d_\text{lat} \sec(\varphi_\text{PD}) \sec(\varphi_\text{RS}+\varphi_\text{PD}) \sin(\varphi_\text{RS}) \nonumber \\
 &\,+ d_\text{long}\left[\cos(\varphi_\text{PD}) \sec(\varphi_\text{RS}+\varphi_\text{PD})-1\right] \rbrace\,.
\label{eq: SC OPD withPD}
\end{eqnarray}
Assuming small rotation angles $\varphi_\text{RS}$, we can Taylor expand this equation and find
\begin{eqnarray}
\text{OPD}_\text{RS}^\text{2D} \approx -d_\text{lat}\, \varphi_\text{RS} + d_\text{long}\, \varphi_\text{RS}^2/2  \,
\label{eq: SC OPD series}
\end{eqnarray}
or, having a tilted photodiode, 
\begin{eqnarray}
\text{OPD}_\text{RS}^\text{2D} \approx 
 -d_\text{lat}\, \varphi_\text{RS} 
 + d_\text{long} \left(\varphi_\text{RS}^2/2+\varphi_\text{RS}\varphi_\text{PD}\right) .
\label{eq: SC OPD with PD series}
\end{eqnarray}
Hence, the lateral distance between the beam axis and the centre of rotation makes a linear TTL coupling and the longitudinal distance adds a second-order coupling term. 
This means that particularly lateral displacements between the centre of rotation and the beam axis should be avoided. Ideally, the centre of rotation should lie in the measurement beam incidence point on the photodiode. In that case, the geometric TTL coupling described here would vanish entirely. 

The equations presented for the receiving system have so far been derived for angular jitter coupling, only. Contrary to the setup with a jittering mirror, lateral jitter does not couple into the geometric length readout in the receiver case. 
This is due to the definition of the lateral jitter axis, which is an axis fixed to the receiving system. In Fig.~\ref{fig:geoSCTTL} it is parallel to the detector surface. Thus, the measurement beam would walk along the detector surface, but its length would stay unchanged.
However, if the photodiode happens to be misaligned by an angle $\varphi_\text{PD}$ with respect to the receiver coordinate system, we find for lateral receiving system jitter $y_\text{RS}$ also linear TTL coupling:
\begin{eqnarray}
\text{OPD}_\text{RS}^\text{2D} &= y_\text{RS}\,\frac{\sin(\varphi_\text{PD})}{\cos(\varphi_\text{RS}+\varphi_\text{PD})} 
\label{eq:RS_lateral}\\
&\approx y_\text{RS}\, \varphi_\text{PD} \,.
\label{eq:RS_lateral_series}
\end{eqnarray}

\subsection{Comparison of mirror and system rotation}
\label{sec: compare_SC_TM}
We now want to compare the geometric TTL coupling in the case of a mirror rotation featuring a beam of normal incidence at the mirror (Eq.~\eqref{eq: piston sec order simple}) with the case of a system rotation (Eq.~\eqref{eq: SC OPD series}). The two equations are identical, except for a factor of two. 
This is not surprising since in the first case, the rotation of the system effectively causes the detector surface to move into (or out of) the beam, while in the system rotation case, the mirror surface moves into (or out of) the beam in the very same way. 
The factor of two is then coming from the fact that the mirror surface reflects the beam, such that the beam propagates the shortened (or additional) length twice. However, in the case of the system rotation, the detector absorbs the beam, such that the path length change is not doubled.

\subsection{Extension for broader set of applications}
\label{sec:geom:extensions}
We will now extend the geometric TTL equations introduced above for a wider set of applications. We will start with the effect of plane-parallel transmissive components along the beam path in \cref{sec: geom tc}. Then, we discuss in \cref{sec:par:exp_in_air} how the equations need to be adapted if the setup is located in a medium other than vacuum. In \cref{sec:Nom-disp} we describe the effect of initial misalignment and then discuss in \cref{sec:PD-angle} the relevance of the photodiode angle $\varphi_\text{PD}$. Finally, in \cref{sec: geom3D} we extend the equations for jitter in three-dimensional space.

\subsubsection{Additional plane-parallel transmissive components}
\label{sec: geom tc}
\begin{figure}
	\centering
	\includegraphics[width=0.75\columnwidth]{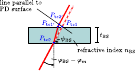}
	\caption{Optical path length changes due to a transmissive optical component: Instead of propagating the distance $\vert \vec{p}_\mathrm{tc1}-\vec{p}_\mathrm{tc0} \vert $ through vacuum (dashed path) the beam propagates from $P_\mathrm{tc0}$ to $P_\mathrm{tc1'}$ through the medium of the component and additionally from $P_\mathrm{tc1'}$ to $P_\mathrm{tc2'}$ through vacuum (solid path). The beam direction within the medium is defined by the angle $\psi_\text{BS}=\arcsin(\sin(\varphi_{\text{BS},i}-\varphi_m)/n_{\text{BS},i})$, where $\varphi_\text{BS}$ defines the alignment of the component with respect to the nominal beam and $\varphi_m$ the additional tilt of the beam, i.e. $\varphi_m\rightarrow2\varphi$ in the case of a mirror rotation and $\varphi_m\rightarrow-\varphi_\text{RS}$ in the case of a rotating receiver.}
	\label{fig: transmissive component}
\end{figure}

For a more complete theory, we introduce transmissive optical components  (tc) with plane-parallel surfaces, e.g. optical windows, beam splitters or beam combiners, as shown in Fig.~\ref{fig: transmissive component}. 
For receiver jitter, we generally need to consider all components along the beam path.
However, in the case of reflection at a jittering mirror, we can ignore all transmissive and reflective components prior to the mirror because these change only the initial conditions (i.e. incidence angle $\beta$ and incidence point $P_2$) of the reference case. 
Their effect can therefore be accounted for by a simple adjustment of the parameters $\beta$ and $d_\text{lat}$, which does not need to be particularly modelled. On the contrary, components along the beam path after the reflection from the jittering mirror directly affect the TTL coupling. 

For all considered components, we assume a thickness $t_{\text{BS},i}$, a refractive index $n_{\text{BS},i}$ and that their surface normal in a two-dimensional setup is rotated by an angle $\varphi_{\text{BS},i}$ against the beam direction of the initial beam. 
Hence, the beams hit the transmissive component under an angle $\varphi_\text{BS}-\varphi_m$, where $\varphi_m$ defines the orientation of the beam, i.e. $\varphi_m\rightarrow2\varphi$ in the case of a mirror rotation and $\varphi_m\rightarrow-\varphi_\text{RS}$ in the case of a rotating system.
While the beam travels with the speed of light through vacuum, it slows down when entering the component $i$. Thus its path length within the component scales with $n_{\text{BS},i}$ and seen by the photodiode the point of reflection seems to be further away. 

Additional to the path length scaling, the direction of the beam changes when it enters a component as shown in Fig.~\ref{fig: transmissive component}. The new beam direction relative to the surface normal vector of the component $i$ is given by the angle of refraction  
\begin{eqnarray}
\psi_\text{BS} = \arcsin\left(\frac{\sin(\varphi_{\text{BS},i}-\varphi_m)}{n_{\text{BS},i}}\right) \, .
\end{eqnarray}
When leaving the component, the beam continues propagating in its original direction. 
The additional optical path length by each transmissive component is then given by 
\begin{equation}
\text{OPL}_\text{tc} = n_{\text{BS}}\, \vert \vec{p}_\mathrm{tc1'}-\vec{p}_\mathrm{tc0}\vert + \vert\vec{p}_\mathrm{tc2'}-\vec{p}_\mathrm{tc1'}\vert  - \vert \vec{p}_\mathrm{tc1}-\vec{p}_\mathrm{tc0}\vert \,,
\label{eq:OPL1tc}
\end{equation}
which is the new path length inside the respective component $\vert \vec{p}_\mathrm{tc1'}-\vec{p}_\mathrm{tc0}\vert$ scaled with its refractive index $n_\mathrm{BS}$ plus the distance between the point $P_\mathrm{tc1'}$ where the beam leaves the component and the intersection point $P_\mathrm{tc2'}$ between the new beam path and the line parallel to the detector (which also intersect with the original exit point $P_\mathrm{tc1}$), i.e. $\vert \vec{p}_\mathrm{tc2'}-\vec{p}_\mathrm{tc1'}\vert$.
From that we have to subtract the original path length $\vert \vec{p}_\mathrm{tc1}-\vec{p}_\mathrm{tc0}\vert$.
We can evaluate Eq.~\eqref{eq:OPL1tc} for $N$ transmissive components 
\begin{eqnarray}
\text{OPL}_\text{tc}^\text{2D} &= \sum_{i=1}^N n_{\text{BS},i} & \,\vert \vec{p}_{\mathrm{tc1'},i}-\vec{p}_{\mathrm{tc0},i}\vert - \vert \vec{p}_{\mathrm{tc1},i}-\vec{p}_{\mathrm{tc0},i}\vert + \vert\vec{p}_{\mathrm{tc2'},i}-\vec{p}_{\mathrm{tc1'},i}\vert
\\
 &= \sum_{i=1}^N \, t_{\text{BS},i} &\Big\lbrace \, n_{\text{BS},i}^2 \left[n_{\text{BS},i}^2-\sin^2(\varphi_m-\varphi_{\text{BS},i})\right]^{-1/2} %
 - \sec(\varphi_m-\varphi_{\text{BS},i})  \nonumber\\
 & &\ + \sin(\varphi_{\text{BS},i}-\varphi_{\text{PD}})\, \sec(\varphi_m-\varphi_{\text{PD}}) \nonumber\\
 & &\ \: \cdot \Big[\sin(\varphi_m-\varphi_{\text{BS},i})\ \left[n_{\text{BS},i}^2-\sin^2(\varphi_m-\varphi_{\text{BS},i})\right]^{-1/2} \nonumber\\
 & &\quad\ - \tan(\varphi_m-\varphi_{\text{BS},i}) \Big] \Big\rbrace
\label{eq: tc}
\\
\text{OPD}_\text{tc}^\text{2D} & \approx \sum_{i=1}^N \frac{t_{\text{BS},i}}{2} &\Big\lbrace 
 n_{\text{BS},i}^2 \cos^2(\varphi_{\text{BS},i}) \left[n_{\text{BS},i}^2-\sin^2(\varphi_{\text{BS},i})\right]^{-3/2} \nonumber\\
 & &\ - \sin^2(\varphi_{\text{BS},i}) \left[n_{\text{BS},i}^2-\sin^2(\varphi_{\text{BS},i})\right]^{-1/2} \nonumber\\
 & &\ - \cos(\varphi_{\text{BS},i}) \Big\rbrace
 \left(\varphi^2_m - 2\varphi_m\, \varphi_{\text{PD}}\right) \,.
\label{eq: tc sec order}
\end{eqnarray}
According to Eq.~\eqref{eq: tc sec order} the optical cross-talk due to transmissive components is a second-order effect. For a few thin components and rather small angles $\varphi_{\text{BS},i}$ we find it to be smaller than the lever arm effect. Nevertheless, it should not be neglected when high accuracy is required. \\
The geometric TTL effects described above and the OPL change due to transmissive components, as it was defined in this section, are independent of one another. 
We, therefore, get the full geometric coupling by adding OPD$_\text{tc}$ to the previously found TTL effects, namely
\begin{eqnarray}
\text{OPD}_\text{MRT} &= \text{OPD}_\text{lever} + \text{OPD}_\text{piston} + \text{OPD}_\text{tc} \, , 
\label{eq: geom full}\\ 
\text{OPD}_\text{RST} &= \text{OPD}_\text{RS} + \text{OPD}_\text{tc} \, .
\label{eq: geom SC full}
\end{eqnarray} 

\subsubsection{Experiments in air or other surrounding medium}
\label{sec:par:exp_in_air}
The above equations of the lever arm and piston effect, $\text{OPL}_\text{lever}$ and $\text{OPL}_\text{piston}$, as well as the analytic description for rotating systems, describe the changes in the OPL while assuming that the beams propagate through vacuum. If this is not the case, we need to multiply all found OPLs in Sec.~\ref{sec: SCgeom} and Sec.~\ref{sec: geom} by the medium's refractive index $n$. 
For transmissive components, we replace $n_{\text{BS},i}$ in equations \eqref{eq: tc} and \eqref{eq: tc sec order} by $n_{\text{BS},i}/n$ and multiply the full path length change by $n$.
This additional factor originates from the lower speed of light in a medium than in a vacuum resulting in an increase of the light travelling time in the medium, which then results in the larger phase change. Therefore, the beam paths appear longer when detected at the photodiode unless the refractive index is known and accounted for in the conversion of the phase into the LPS signal.

\subsubsection{Additional misalignment}
\label{sec:Nom-disp}
In sections \ref{sec: geom} and \ref{sec: geom tc} we compared the initial beam path length with the path length of a beam that got reflected at a misaligned mirror.
In experiments, we often face the case that the mirror $M_\text{tilt}$ is nominally misaligned and we are interested in path length changes to this nominal case. This can be easily implemented into our formulas by substituting $\varphi\rightarrow\varphi^\prime+\varphi_0$ and $d_\text{lat}\rightarrow d_\text{lat}^\prime+y_0$, where $\varphi_0$ denotes the nominal tilt and $y_0$ the nominal lateral shift of the mirror. All OPDs are then defined as the difference of the OPL for an arbitrary angle $\varphi^\prime$ minus the OPL at a nominal angle, i.e. at  $\varphi^\prime=0$:
\begin{equation}
	\text{OPD} = \text{OPL}(\varphi^\prime)- \text{OPL}(\varphi^\prime=0) \;.
\end{equation}
Angular misalignment can likewise happen in the case of a rotating system (see  Sec.~\ref{sec: SCgeom}). Due to experimental tolerances, the system is likely slightly misaligned with respect to the incoming beam. The equations in Sec.~\ref{sec: SCgeom} can then be adapted analogously.

\subsubsection{Relevance of the photodiode angle}
\label{sec:PD-angle}
The OPD depends on the photodiode angle via
\begin{eqnarray}
&\text{OPD}_\text{MRT}(\varphi_\text{PD})-\text{OPD}_\text{MRT}(\varphi_\text{PD}=0) \nonumber\\
= &\Bigg\lbrace  
 \sum_{i=1}^N t_{\text{BS},i} \Big\lbrace 
 n_{\text{BS},i}^2 \cos^2(\varphi_{\text{BS},i}) \left[(n_{\text{BS},i}^2-\sin^2(\varphi_{\text{BS},i})\right]^{-3/2} \nonumber\\
 &\qquad\qquad\ - \sin^2(\varphi_{\text{BS},i}) \left[n_{\text{BS},i}^2-\sin^2(\varphi_{\text{BS},i})\right]^{-1/2} 
 - \cos(\varphi_{\text{BS},i}) \Big\rbrace \nonumber\\
 &\ + d_\text{lever} + d_\text{lat}\,\sin(\beta) 
 \Bigg\rbrace 
 \left(- 2\varphi\, \varphi_{\text{PD}}\right)  \,, \\
&\text{OPD}_\text{RST}(\varphi_\text{PD})-\text{OPD}_\text{RST}(\varphi_\text{PD}=0) \nonumber\\
= &\Bigg\lbrace  
 \sum_{i=1}^N t_{\text{BS},i} \Big\lbrace 
 n_{\text{BS},i}^2 \cos^2(\varphi_{\text{BS},i}) \left[(n_{\text{BS},i}^2-\sin^2(\varphi_{\text{BS},i})\right]^{-3/2} \nonumber\\
 &\qquad\qquad\ - \sin^2(\varphi_{\text{BS},i}) \left[n_{\text{BS},i}^2-\sin^2(\varphi_{\text{BS},i})\right]^{-1/2} 
 - \cos(\varphi_{\text{BS},i}) \Big\rbrace \nonumber\\
 &\ + d_\text{long}
 \Bigg\rbrace 
 \left(\varphi_\text{RS}\, \varphi_{\text{PD}}\right)  \,.
\end{eqnarray}

This is natural due to the geometric nature of the OPD. However, the tilt of the detector affects how the wavefronts of both beams, the measurement and reference beam, are being detected. We will, therefore, show in \cite{NG21} that the non-geometric effects contribute further terms that cancel the detector orientation from the total interferometric readout. All terms containing $\varphi_\text{PD}$ should then be handled with care: they do exist in the OPD but do not affect the total interferometric readout. It can be an advantage to neglect the photodiode angle $\varphi_\text{PD}$=0 in simulations, even in cases where the corresponding experiment comprises large angles.

\subsubsection{Three-dimensional case}
\label{sec: geom3D}
So far, we only considered a two-dimensional setup, which can then be applied in 3D setups for changes in a specific plane. If angular jitter occurs in two orthogonal planes, one can simply add up the effect of the individual planes, provided that we have beams with normal incidence and the jitter is small and therefore describable by a first-order series expansion. If the jitter causing the TTL coupling has a larger amplitude, effects occurring in one plane also affect the orthogonal plane.
This can, for instance, be seen in the lever arm effect, see Eq.~\eqref{eq: piston sec order}. There, a rotation in the orthogonal plane would affect the length of the now three-dimensional lever arm $d_\text{lever}$, which results in a product term of the involved angles, i.e. $\varphi \cdot\eta$.

As illustrated in \cref{fig:geoTTL} and \cref{fig:geoSCTTL}, all equations given so far were describing motion and alignment in the $xy$-plane. The angle $\varphi$ (or $\varphi_\text{RS}$ respectively) is, therefore, a yaw angle originating from a rotation around the $z$-axis. The orthogonal pitch rotation $\eta$ (or $\eta_\text{RS}$) is then a rotation around the $y$-axis. The cross-coupling between yaw and pitch makes the equations significantly more complex. We, therefore, series expand the expressions for the 3D geometric effects to second-order and find:
\begin{eqnarray}
\text{OPD}_\text{lever}^\text{3D} &= 
d_\text{lever} \Bigg[\frac{ 
2\cos^2(\beta_z)(\varphi^2-\varphi\varphi_\text{PD})
+2\cos^2(\beta_y)(\eta^2-\eta\eta_\text{PD})}
{\cos^2(\beta_y) + \cos^2(\beta_z) \sin^2(\beta_y)} \nonumber\\
&\; +\frac{\sin(2\beta_y)\sin(2\beta_z)\, \varphi\,\eta}
{\cos^2(\beta_y) + \cos^2(\beta_z) \sin^2(\beta_y)}\Bigg]
\label{eq: lever_3D} \\
\text{OPD}_\text{piston}^\text{3D} &=
2 \left(-d_\text{lat}\, \varphi +d_\text{vert}\, \eta \right) \biggg[
\frac{\cos(\beta_y)\cos(\beta_z)}
{\sqrt{\cos^2(\beta_y) + \cos^2(\beta_z) \sin^2(\beta_y)}} \nonumber\\
&\; - \frac{ \sin(\beta_y)\cos(\beta_z)(\varphi-\varphi_\text{PD}) + \sin(\beta_z)\cos(\beta_y)(\eta-\eta_\text{PD}) }
{\sqrt{\cos^2(\beta_y) + \cos^2(\beta_z) \sin^2(\beta_y)}} \biggg]
\nonumber\\
&+ 
d_\text{long}\,\frac{\cos(\beta_y)\cos(\beta_z)\, \left(\varphi^2+\eta^2\right) }{\sqrt{\cos^2(\beta_y) + \cos^2(\beta_z) \sin^2(\beta_y)}}
\label{eq: piston_3D} \\
\text{OPD}_\text{RS}^\text{3D} &= 
- d_\text{lat}\, \varphi_\text{RS}  + d_\text{vert}\, \eta_\text{RS} +d_\text{long}\, \left(\frac{\varphi_\text{RS}^2+\eta_\text{RS}^2}{2}\right) \,,
\label{eq: SC_3D} 
\end{eqnarray}
where $\beta_y, \beta_z$ are the projection-angles of the incoming beam at the mirror into the $xy$- and $xz$-plane, respectively.
The different signs of the $d_\text{lat}$- and $d_\text{vert}$-terms arise from the orientations of the pitch and yaw angles.

Photodiode tilts in the $xz$-plane are indicated by $\eta_\text{PD}$, and the vertical difference between the point of incidence and the point of reflection is given by $d_\text{vert}$. 
Naturally, all three-dimensional equations reduce to their two-dimensional counterparts if the parameters in the respective orthogonal plane are set to zero.

Another special case arises if the nominal beams hit the test mass orthogonally, i.e. $\beta_{y,z}=0$. Here, all cross-plane couplings vanish and the resulting signal equals the sum of the two two-dimensional equations applying the same assumption. We get

\begin{eqnarray}
\text{OPD}_\text{lever}^\text{3D} &=\,
d_\text{lever} \left[ 2(\varphi^2-\varphi\varphi_\text{PD})
+2(\eta^2-\eta\eta_\text{PD}) \right]
\label{eq: lever_3D_beta0} \\
\text{OPD}_\text{piston}^\text{3D} &=\, 
d_\text{long}\left(\varphi^2+\eta^2\right)
+2 \left(-d_\text{lat}\, \varphi +d_\text{vert}\, \eta \right)
\label{eq: piston_3D_beta0}
\end{eqnarray}

\begin{figure}
\centering
	\includegraphics[width=\columnwidth]{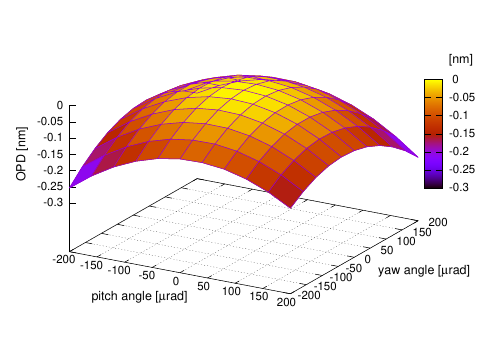}
	\caption{Optical path length difference of a beam, that got tilted in pitch ($\eta_\text{RS}$) and yaw ($\varphi_\text{RS}$), passing a transmissive optical component. We set $t_\text{tc}=10\,$mm, $n_\text{tc}=2$ and rotated the component with respect to the beam by $\varphi_\text{tc}=0.2\,\pi$ and $\eta_\text{tc}=0.13\,\pi$. While the second-order OPD is dominant, the color grid indicates, that the tilt of the transmissive component also adds linear TTL coupling. }
	\label{fig: OPD tc 3D}
\end{figure}

Also, transmissive components affect the beam in all degrees of freedom. In a three-dimensional setup, we can now additionally consider the components to be rotated in yaw by $\varphi_\text{BS}$ and in pitch by $\eta_\text{BS}$. Hence their 3D-equations become very long such that we summarise here only the principle behaviour. In the case of normal incidence ($\varphi_\text{BS}=\eta_\text{BS}=0$), the transmissive components add a second-order effect. For non-normal incidence, an additional linear effect occurs, see Fig.~\ref{fig: OPD tc 3D}. This linear effect is typically small, such that the second-order effects stay dominant. 

\section{Exemplary setups showing limitations of geometrical estimates}
\label{sec: TTL vs geometry}
The cross-coupling mechanisms explained so far originate from variations of the beam propagation axis that can be described geometrically.
In order to treat these effects as introduced in Sec.~\ref{sec: geom}, we did assume that the laser beams are classical rays.
This assumption, however, is often insufficient since it neglects all wavefront properties. We will show in the examples below that the expected TTL cross-talk can significantly deviate from geometrically expected values if the laser light is instead described as fundamental Gaussian mode.

\subsection{Rotation around the centre of curvature}
\label{ex: TTL radius curvature}
\begin{figure}
	\centering
	\includegraphics[width=0.6\columnwidth]{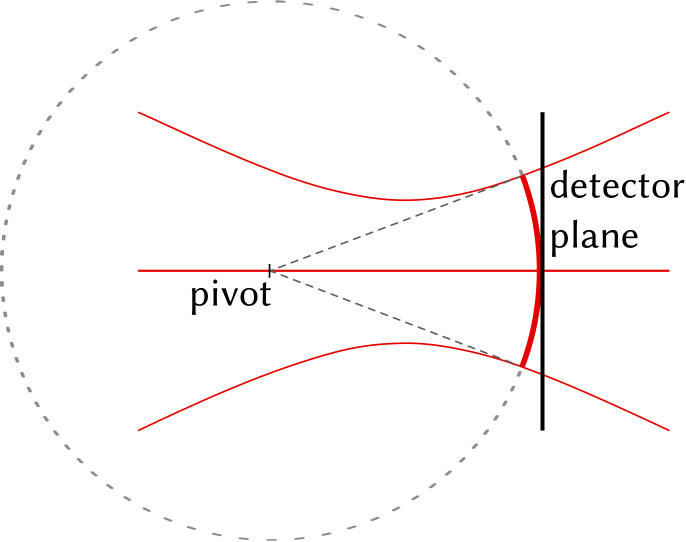}
	\caption{\label{fig: Gaussian curvature sima}A Gaussian beam is tilted around a pivot that coincides with the centre of the wavefront curvature at the detector position. While the beam is tilted, the phase distribution on the detector stays unchanged.}
\end{figure}

Assume the setup previously published in ~\cite{Wanner10PhD, wanner2012methods} and illustrated in Fig.~\ref{fig: Gaussian curvature sima}. There, we consider two identical Gaussian beams, i.e.\ with the same intensity distribution and phase fronts. We rotate one of the beams (measurement beam) with respect to the other around a pivot that has a longitudinal distance to the detector that is equal to the radius of curvature of the wavefront on the detector.
\begin{figure}
	\centering
	\includegraphics[width=0.85\columnwidth]{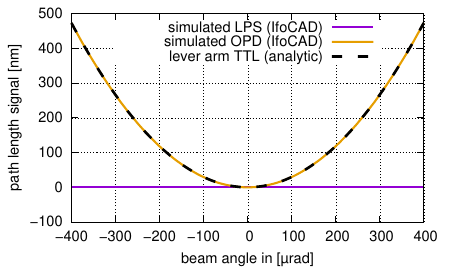}
	\caption{\label{fig: Gaussian curvature simb}Simulated path length signals (i.e.\ LPS and OPD) compared to the expected analytic TTL coupling in the scenario shown in Fig.~\ref{fig: Gaussian curvature sima}, i.e.\ the lever arm coupling if the rotation is induced by a rotating test mass. Hence the beam angle $\varphi_m$ depends on the mirror tilt via $\varphi_m=2\varphi$. The analytically derived TTL coupling (black, dashed) and the simulated OPD (yellow) are non-zero and coincide. However, this lever arm TTL coupling cannot be observed in the simulation (purple) due to the special conditions of this scenario (rotation without change of the wavefront). In this simulation, we assumed identical Gaussian beams with a waist radius $w_0=$1\,mm, distance from waist at detector $z=z_R$, lever arm $d_\text{lever}=2z_R$, single element photodiode with radius 5\,mm.}	
\end{figure}
In this setup, a lever arm between the point of rotation and the detector should lead to significant cross-coupling. %
However, in this scenario, a tilt of the measurement beam does not change the phase distribution on the detector since the wavefront is mapped onto itself during the rotation. That means only the intensity distribution on the detector plane changes during the rotation. If, however, the reference wavefront curvature equals the measurement wavefront curvature, then the phase difference is zero in every detector point such that zero TTL coupling occurs despite the rotation of the beam axis.

To demonstrate that, in fact, no cross-coupling is present in this scenario, we numerically compute the variations of the longitudinal path length sensing (LPS) signal, i.e. the interferometric phase signal processed by the detector converted to units of lengths. In the computation of this signal, not only the OPD but also wavefront and detector properties are being considered (see \cite{wanner2012methods} and \cite{NG21} for details on the numerical computation of this signal). 
This was done using the simulation tool IfoCAD~\cite{wanner2012methods,Kochkina2013}. 
The obtained LPS signal is shown in Fig.~\ref{fig: Gaussian curvature simb}. 
We compare it with the numerically derived OPD, which corresponds to the analytic expression of the lever arm cross-coupling Eq.~\eqref{eq: lever arm}. Unlike the geometric description, the simulated LPS signal shows no cross-coupling, just as expected, demonstrating that the pure geometric description of this scenario is not sufficient.
The resulting TTL coupling can only be correctly estimated when non-geometric effects are considered as well. 

\subsection{Identical fundamental Gaussian beams with centre of rotation on the beam axis}
\label{ex: identical Gaussian beams}
\begin{figure}
	\centering
	\includegraphics[width=0.85\columnwidth]{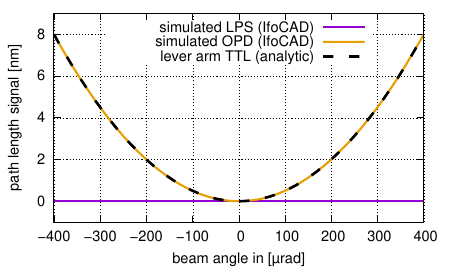}
	\caption{\label{fig: Gaussian vanishing ttl}Simulated path length signal compared to the expected lever arm TTL coupling in the scenario with two equal Gaussian beams and a large single element detector. The beam angle equals two times the mirror tilt. The TTL coupling vanishes due to symmetric phase cancellations. The simulation parameters were: waist radius $w_0=$1\,mm, distance from waist at the detector $z=100\,$mm,  lever arm $d_\text{lever}=100\,$mm, detector radius 100\,mm.}%
\end{figure}
The cross-coupling also vanishes for two identical Gaussian beams with one beam tilted around a pivot that has now an arbitrary longitudinal but no lateral distance to the detector. In this case, we discover a lever arm effect and not necessarily equal wavefronts. 
However, we find no cross-coupling as illustrated in Fig.~\ref{fig: Gaussian vanishing ttl} and introduced in \cite{Schuster2015}. We conclude that a non-geometric cross-coupling mechanism caused by the tilt dependent beam walk on the detector cancels the expected lever arm cross-coupling on a sufficiently large detector. A general analytic description of the non-geometric effect is given in the next section. Additionally, an intuitive approach to understand the cancellation is given in \ref{app: signal cancellation}. \\

\section{Summary geometric TTL coupling and TTL mitigation strategies}
\label{Sec:GeomEffectsSummary+Mitigation}
In this paper, we have introduced different geometric TTL coupling effects and derived equations that describe whether they add primarily linear, quadratic or mixed terms to the OPD. 
All effects are summarised in Tab.~\ref{tab: TTL_mech_overview} for convenience.
Lateral jitter coupling appears as a linear effect only in the case of mirror reflection:
\begin{eqnarray}
	&\text{OPD}_\text{piston,lat}^\text{2D}(d_\text{lat}(t)) &\approx -2d_\text{lat}(t)\,\varphi \,,
	\label{eq:sum_OPD_piston_lat-y} \\
	&\text{OPD}_\text{RS,lat}^\text{2D} &= 0 \,.
	\label{eq:sum_OPD_SC_lat-y} 
\end{eqnarray}
For angular jitter TTL coupling we have the following linear TTL effects:
\begin{eqnarray}
	\text{OPD}_\text{piston,lat}^\text{2D}(\varphi(t)) &\approx -2d_\text{lat}\,\varphi(t) \,,
	\label{eq:sum_OPD_piston_lat-phi} \\
	\text{OPD}_\text{RS,lat}^\text{2D}(\varphi_\text{RS}(t)) &\approx -d_\text{lat}\,\varphi_\text{RS}(t) \,,
	\label{eq:sum_OPD_SC_lat-phi} 
\end{eqnarray}
which are caused by lateral displacements of the centre of rotation with respect to the principle beam axis. Second-order TTL effects are:
\begin{eqnarray}
	&\text{OPD}_\text{lever}^\text{2D} &\approx 2d_\text{lever}\,\varphi^2 \,,
	\label{eq:sum_OPD_lever} \\
	&\text{OPD}_\text{piston,long}^\text{2D} &\approx d_\text{long}\,\varphi^2  \,,
	\label{eq:sum_OPD_piston_long-phi} \\
	&\text{OPD}_\text{RS,long}^\text{2D} &\approx \frac{1}{2}d_\text{long}\,\varphi_\text{RS}^2 \;,
	\label{eq:sum_OPD_SC_long-phi}
\end{eqnarray}
which all originate from longitudinal displacements of the centre of rotation with respect to either the beam incidence point on the reflecting component ($d_\text{long}$), or the photodiode ($d_\text{long}$, $d_\text{lever}$). An additional second-order TTL effect is caused by transmissive components:
\begin{eqnarray}
	\text{OPD}_\text{tc}^\text{2D} & \approx \sum_{i=1}^N \frac{t_{\text{BS},i}}{2} &\Big\lbrace 
 n_{\text{BS},i}^2 \cos^2(\varphi_{\text{BS},i}) \left[n_{\text{BS},i}^2-\sin^2(\varphi_{\text{BS},i})\right]^{-3/2} \nonumber\\
 & &\ - \sin^2(\varphi_{\text{BS},i}) \left[n_{\text{BS},i}^2-\sin^2(\varphi_{\text{BS},i})\right]^{-1/2} \nonumber\\
 & &\ - \cos(\varphi_{\text{BS},i}) \Big\rbrace\,
 \varphi^2_m \,,
	\label{eq:sum_OPD_tc}
\end{eqnarray}
with $\varphi_m=2\varphi$ for components between the mirror and the detector and $\varphi_m=-\varphi_\text{RS}$ for components along the beam path inside a receiving system with angular jitter.
The knowledge about these first- and second-order couplings can be used to mitigate the geometric effects by smart design choices, for instance, by aligning the beam axis to the centre of rotation ($d_\text{lat}=0$) or by imaging the centre of rotation to the photodiode centre, resulting in $d_\text{long}=d_\text{lever}=0$. 
Lateral jitter coupling is minimised in the case of a mirror reflection by an angular adjustment to make $\varphi=0$. %
Moreover, the given equations can be used to minimise the total TTL coupling (see Eq.~\eqref{eq:TTL_ng+g}). In that case, the geometric TTL is manipulated to counteract the non-geometric TTL effects (LPS$_\text{ng,TTL}$). These non-geometric effects are more difficult to describe or mitigate, which will be discussed in detail in our follow-up paper \cite{NG21}.
For such mitigation, the total TTL of the system is measured, i.e.\ the sum of all geometric and non-geometric effects. The total linear effects can then be minimised by adjusting the lateral alignment parameter $d_\text{lat}$, for instance, by adjusting the beam axis.
The sum of all second-order effects is in the next step minimised by adjusting either of the longitudinal parameters ($d_\text{long}, d_\text{lever}$), for instance, by displacing the photodiode longitudinally. Such mitigation was experimentally used, for instance, in \cite{Troebs2018, Chwalla2020}. \\
It should be noted, that this type of TTL noise suppression is only successful if the applied changes affect the geometric TTL coupling more than the non-geometric TTL effects. This is discussed further in \cite{NG21}.

\fulltable{	\label{tab: TTL_mech_overview}Overview of the different geometric cross-coupling mechanisms for a beam with normal incidence and no photodiode tilt, i.e. $\beta=\varphi_\text{PD}=0$. For each effect we give a short description and the general behaviour (approximated), like linear, quadratic or mixed with respect to the tilt angle.}
	\renewcommand{\arraystretch}{1.5}
\begin{tabular*}{1\textwidth}{@{\extracolsep{\fill}}>{\raggedright}l| >{\raggedright}p{0.145\textwidth} l >{\centering}p{0.105\textwidth} c c p{0.385\textwidth}}
	\br
	& Cross-coupling mechanism  &  Name  &  General behaviour  &  Eq.  &  Sec.  &  Description
	\\
	\hline
	\parbox[t]{3mm}{\multirow{3}{*}{\rotatebox[origin=c]{90}{lateral}}}
	& piston effect & $\text{OPD}_\text{piston}$ & linear & \eqref{eq:sum_OPD_piston_lat-y} & \ref{sec: geom} & If the mirror is initially rotated, further lateral shifts generate TTL coupling. \\
	& receiver jitter & $\text{OPD}_\text{RS}$ & zero & \eqref{eq:sum_OPD_SC_lat-y} & \ref{sec: SCgeom} & 
	Lateral jitter of the receiver does not cause geometric TTL coupling.
	\\\hline
	\parbox[t]{3mm}{\multirow{10}{*}{\rotatebox[origin=c]{90}{angular}}}
	& lever arm effect & $\text{OPD}_\text{lever}$ & quadratic & \eqref{eq:sum_OPD_lever}
	& \ref{sec: geom} & Longitudinal offsets between rotation point and detector lead to variations in the propagation distance to the detector.
	\\
	& \multirow{3}{*}{}piston effect & $\text{OPD}_\text{piston}$ & & & \ref{sec: geom} & \multirow{2.5}{0.385\textwidth}
	{\justifying Offsets between the points of rotation and reflection at a component's surface generate movement of this surface.} \\ [-3pt]
	& $-$ longitudinal & & quadratic & \eqref{eq:sum_OPD_piston_long-phi} & &
	\\ [-3pt]
	& $-$ lateral & & linear & \eqref{eq:sum_OPD_piston_lat-phi} & &
	\\
	& \multirow{3}{*}{}setup jitter & $\text{OPD}_\text{RS}$ & & & \ref{sec: SCgeom} & \multirow{3.15}{0.385\textwidth}
	{\justifying Offsets between the points of rotation and detection generate movement of the detector surface relative to the beam axis.} \\ [-3pt]
	& $-$ longit. CoR & & quadratic & \eqref{eq:sum_OPD_SC_long-phi} & &
	\\ [-3pt]
	& $-$ lateral CoR & & linear & \eqref{eq:sum_OPD_SC_lat-phi} & & \\
	& transmissive components & $\text{OPD}_\text{tc}$ & quadratic & \eqref{eq:sum_OPD_tc} & \ref{sec: geom tc} & Transmissive optical components between the mirror and the detector induce a variation of the beams' optical path length and a beam displacement on the detectors surface.
	\\
	\br
\end{tabular*}
\endfulltable

\section{Conclusion}
\label{sec: summary}
Within this paper, we have discussed how and why angular and lateral jitter causes phase noise in laser interferometers. 
This tilt-to-length (TTL) noise is often dominated by length changes of the beam axis, i.e.\ geometric effects, which are the focus of this paper. 
We have categorised different coupling mechanisms and analytically derived their contribution to the interferometric path length signal from trigonometric relationships.

We have introduced TTL coupling noise for two different cases: jitter of a mirror within the interferometer and jitter of the receiving system. Though the described effects are independent of the context, they apply, for instance, for the LISA mission and other space interferometers. In that context, mirror jitter corresponds to the case observed in test mass interferometers, while receiver jitter corresponds to long arm interferometers and effectively describes satellite jitter relative to the received wavefront. 

Besides the geometric TTL contributions, there exists non-geometric TTL coupling originating from the wavefront properties of the interfering beams as well as from detector characteristics. We have presented two examples demonstrating the relevance of the non-geometric TTL contributions. However, these non-geometric effects are significantly harder to describe than their geometric counterparts and are the subject of a follow-up paper \cite{NG21}. 

The TTL coupling within a system will always be the sum of all present non-geometric and geometric TTL contributions. 
Nevertheless, knowing the properties of the geometric TTL contributions alone can still help suppress TTL noise in precision laser interferometers in two different ways.  

On the one hand, the geometric effects can a priori be suppressed by smart design choices. For the suppression of angular jitter coupling, it is, for instance, advisable to particularly match the lateral position of the centre of rotation and the reflection point on the jittering mirror (or test mass) with high precision because the residual offset causes a linear effect (Eq.~\eqref{eq:sum_OPD_piston_lat-phi}). %
Using imaging optics as described in \cite{Troebs2018, Chwalla2020} can significantly reduce the TTL coupling. Such imaging systems image the centre of rotation to the photodiode centre, which results in $d_\text{lever} = d_\text{long}=0$ and minimise the lever arm effect (Eq.~\eqref{eq:sum_OPD_lever} or Eq.~\eqref{eq:sum_OPD_SC_long-phi}, respectively). 
Also, the TTL coupling from lateral jitter can be reduced by respective design choices. Here, angular misalignments of the mirror %
need to be minimised (Eq.~\eqref{eq:sum_OPD_piston_lat-y}). %
Therefore, the effects described here allow for a TTL coupling suppression by smartly designing the interferometric layout. 

On the other hand, the known geometrical TTL mechanism can be used to counteract the measured TTL noise in an experiment. From experimental data, one cannot necessarily tell apart the geometric and non-geometric effects. Instead, the total TTL noise is being measured. In some cases, one can then use \cref{tab: TTL_mech_overview} to counteract the total measured TTL. For instance, one can laterally adjust the reflection point on the jittering mirror to balance the sum of all existent first-order effects. This could potentially mean that the lateral piston is then intentionally being increased.
Likewise, if the sum of all second-order effects is sufficiently small and if space in the experimental layout permits, one can, for instance, shift the detector longitudinally (altering the magnitude of the lever arm effect) in order to minimise the observed second-order TTL noise.

In summary, we have introduced and analytically described a variety of geometric TTL coupling effects. These can add significant noise to the signal of laser interferometers. The mechanisms described here allow the suppression of TTL noise by a dedicated design optimisation, as well as in the experimental setup by a dedicated realignment to counteract the observed TTL noise.

\ack
We thank Gerhard Heinzel for valuable discussions.
This work was made possible by funds of both the Deutsche Forschungsgemeinschaft (DFG) and the German Space Agency, DLR.
We gratefully acknowledge the Deutsche Forschungsgemeinschaft (DFG) for funding the Sonderforschungsbereich (SFB 1128: geo-Q) ``Relativistic Geodesy and Gravimetry with Quantum Sensors'', project A05 and all work contributions to this paper made by Sönke Schuster. 
Furthermore, we acknowledge DFG for funding the Clusters of Excellence PhoenixD (EXC 2122, Project ID 390833453) and QuantumFrontiers (EXC 2123, Project ID 390837967). 
Likewise, we gratefully acknowledge the German Space Agency, DLR and support by the Federal Ministry for Economic Affairs and Energy based on a resolution of the German Bundestag (FKZ 50OQ1801). 
Finally, we would like to acknowledge the Max Planck Society (MPG) for supporting the framework LEGACY on low-frequency gravitational wave astronomy, a cooperation between the Chinese Academy of Sciences (CAS) and the MPG (M.IF.A.QOP18098).
\appendix

\section{Signal cancellation for two identical Gaussian beams}
\label{app: signal cancellation}
In Sec~\ref{ex: identical Gaussian beams}, we have shown a case where the total interferometric longitudinal path length sensing (LPS) signal vanishes despite a considerable OPD response. This setup consisted of two identical Gaussian beams, with one rotating around a pivot located at an arbitrary point along the beam axis and a large single element photodiode that detects both incident beams without clipping. 
In this section, we show in a small angle approximation that this LPS signal vanishes due to the symmetry of the interfering wavefronts on the detector surface. We do this in two steps: In \ref{sec:Appendix_amplitude_symmetry}, we show the symmetry of the amplitude profile of the interference pattern, in \ref{sec:Appendix_phase_symmetry} the symmetry of the phase difference of the interfering beams. 
We thereby show that for every point $P_1$ on the detector, there exists a point $P_2$ which is found by mirroring $P_1$ on the symmetry plane. 
In $P_2$, the interference pattern has the same amplitude as in $P_1$, but an inverse phase. 
The LPS signal then vanishes because it is a signal that is integrated over all detector points.

\subsection{Symmetry of the amplitude profile} 
\label{sec:Appendix_amplitude_symmetry}
In the considered setup, we rotate the measurement beam around a point along its propagation axis, while the reference beam is not tilted ($\varphi_r=0$). Let the distance between the point of rotation and the point of incidence, i.e. the point where the ray describing the beam axis impinges on the detector, be given by $d_\text{lever}$. For $d_\text{lever}\neq 0$, the point of incidence gets shifted along the photodiode surface when rotating the beam by $\varphi_m$. Thus the measurement beam's new point of incidence varies by 
\begin{eqnarray}
x_{im} \approx -d_\text{lever} \,\varphi_m \, ,
\end{eqnarray}
while the reference beam stays constant ($x_{ir}=0$).
The amplitude profile of the interference pattern $A(x,y)$, defined according to \cite{Wanner2014} by
\begin{eqnarray}
A(x,y)=\frac{1}{w(z_m)w(z_r)}\exp\left(\frac{-r_m^2(x,y)}{w^2(z_m)}\right)\exp\left(\frac{-r_r^2(x,y)}{w^2(z_r)}\right) \, ,
\label{eq:apx_amplitude}
\end{eqnarray}
is then symmetric to the plane defined by $x=x_c$. 
Here, $w(z_{m,r})$ defines the spot radius of the measurement and reference beam on the detector surface, and $r_{m,r}$ is the cylindrical coordinate of the respective beam.
Furthermore, $x_c$ is the mean value between the $x$-coordinates of the two incidence points
\begin{equation}
x_c = \frac{x_{im}-x_{ir}}{2}  \approx -\frac{1}{2} d_\text{lever}\, \varphi_m\;.
\label{eq:x_c}
\end{equation}
The cylindrical coordinates of the beams when impinging on the detector are given by
\begin{eqnarray}
r_m^2 &=(x-x_{im})^2\cos^2(\varphi_m) + y^2 \,,
\label{eq: rep rm}\\
r_r^2 &= x^2 + y^2 \, .
\label{eq: rep rr}
\end{eqnarray}
The coordinate transformation, performed here on the $x$ and $y$ coordinates, naturally also needs to be performed on the $z$-coordinate \cite{Wanner2014}, resulting in
\begin{eqnarray}
	z_m &\approx z + \frac{1}{2} d_\text{lever} \varphi_m^2 - (x - x_{im}) \,\varphi_m 	\label{eq: rep zm}  \\
	z_r &= z = \text{const}\, .
\end{eqnarray}

Substituting accordingly $z_{m,r}$, $r_{m,r}$ in Eq.~\eqref{eq:apx_amplitude} reveals the symmetry:
\begin{eqnarray}
A(-\frac{1}{2}\, d_\text{lever}+x,y) 
= A(-\frac{1}{2}\, d_\text{lever}-x,y) \,.
\end{eqnarray} 
\subsection{Symmetry of the phase difference and cancellation on the detector} 
\label{sec:Appendix_phase_symmetry}
\begin{figure}
\centering
	\includegraphics[width=\columnwidth]{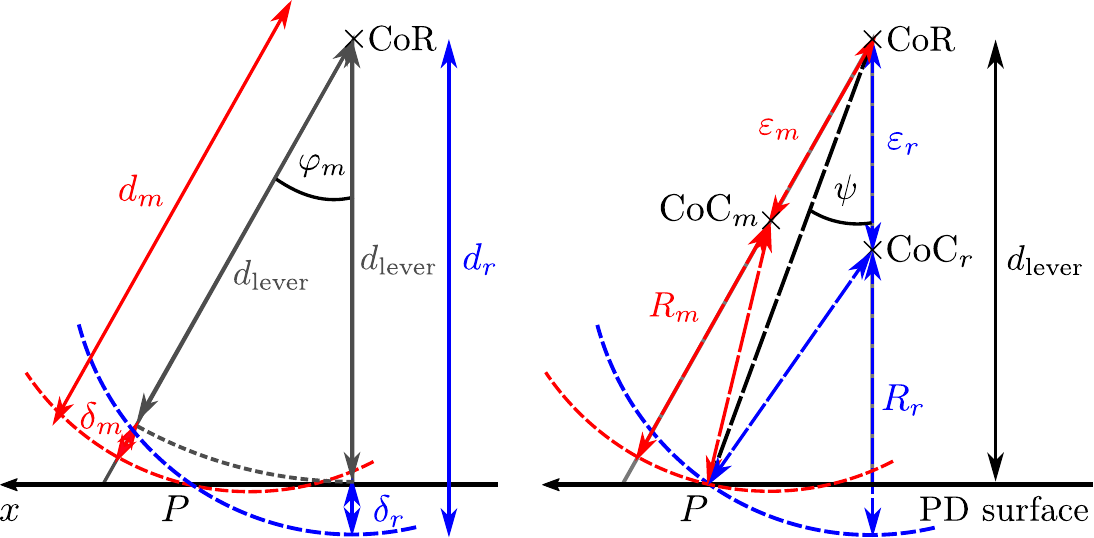}
	\caption{\label{fig: compute vanishing ttl - explain2} Properties of a tilted measurement and a non-tilted reference beam whose wavefronts are intersecting a point $P$ at the detector surface. The paths of the beam rays are described by the solid grey lines. The measurement beam got rotated by an angle $\varphi_m$ (negative in this figure) around the centre of rotation (CoR). 
	Left figure: The beams propagate the distances $d_{m,r}$ from the CoR until their wavefronts hit the point $P$. This distance differs from the $d_\text{lever}$, i.e. the distance between the CoR and the detector, by $\delta_{m,r}$. 
	Right figure: The propagation distance of the beams can also be defined via their radii of curvature $R_{m,r}$. We additionally define the auxiliary lengths $\varepsilon_{m,r}$ as the distances between the CoR and the beam's centres of wavefront curvature (CoC$_{m,r}$). If $R_{m,r}$ are greater than $d_\text{lever}$, the lengths $\varepsilon_{m,r}$ are negative. For every investigated point $P$, there can be found an angle $\psi$ defining its position in relation to the reference beam axis.}
\end{figure}

Next, we examine the local phase differences. We will show that this phase difference is zero at the centre between the points of incidence, while its absolute value increases and decreases symmetrically with increasing distance from this point.
Since the phase relates to a length $l$ via $\phi=2\pi\,l/\lambda$, where $\lambda$ is the wavelength of the beam, we can analogously investigate length changes here.
Therefore, we study the beams' properties and geometric relationships further.
For simplicity, we can reduce our analysis to the plane in which the rotation of the measurement beam gets applied. The phase differences in an orthogonal plane, i.e. for a constant $x$-value, are in a small angle approximation constant.
First, we investigate the Gaussian beam properties. We know that Gaussian beams provide the same phase along their wavefronts. For our analysis, we propagate the beams until their wavefronts hit the point of interest $P$ at the detector, see Fig.~\ref{fig: compute vanishing ttl - explain2}. 
\begin{figure}
\centering
	\includegraphics[width=\columnwidth]{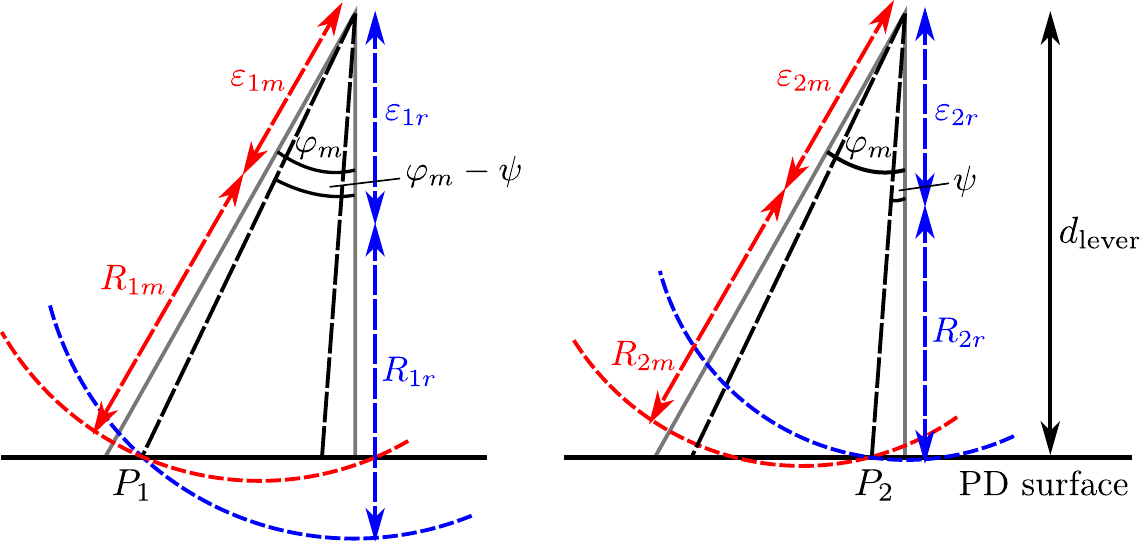}
	\caption{\label{fig: compute vanishing ttl} 
	Investigation of the two beams' local phase differences in the two points $P_1$ (left figure) and $P_2$ (right figure). In both figures, the beam rays are shown as solid grey lines, where the non-tilted right line belongs to the reference and the left line belongs to the measurements beam. $P_1$ and $P_2$ lie symmetrically around the centre between the beams' points of incidence. The points are defined via an angle $\psi$ with respect to the beams' axes. 
	The respective radii of curvature are denoted with $R_{i;m,r}$ and likewise the auxiliary lengths are tagged $\varepsilon_{i;m,r}$.}
\end{figure}
\begin{figure}
\centering
	\includegraphics[width=0.58\columnwidth]{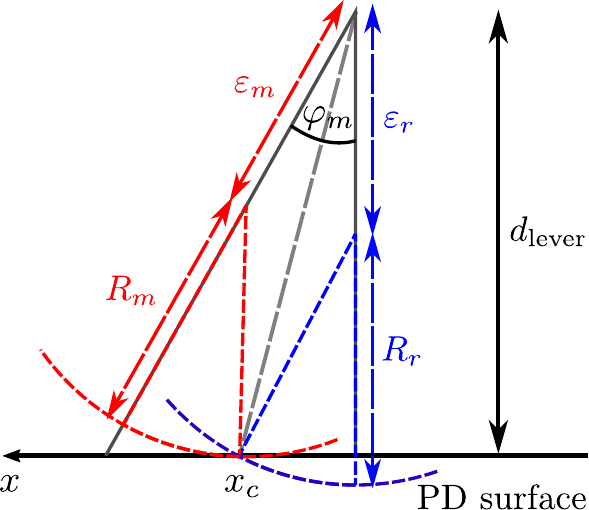}
	\caption{\label{fig: compute vanishing ttl - explain} Investigation of the two beams' local phase differences in the centre between their points of incidence, i.e. for $x=x_c$.
	The measurement (red) and the reference beam (blue) are symmetric with respect to the bisecting line from the centre of rotation to $x_c$.
	Therefore, it is $R_m=R_r$ and $\varepsilon_m=\varepsilon_r$.}
\end{figure}
The beams have then propagated a distance $d_{m,r}$. %
We compare this distance with the lever arm $d_\text{lever}$, which is the distance the beam axis propagates from the CoR to the diode if $\varphi_m=0$. 
We can then show the symmetry of the wavefront using the difference $d_{m,r}-d_\text{lever} =\delta_{m,r}$.
This distance $\delta_{m,r}$ can also be expressed as function of the radii of curvature $R_{m,r}$ and the auxiliary lengths $\varepsilon_{m,r}$:
\begin{eqnarray}
 \delta_{m,r} =R_{m,r} + \varepsilon_{m,r}  - d_\text{lever} \,,
\label{eq: system of eq epsilon}
\end{eqnarray}
as visible in Fig.~\ref{fig: compute vanishing ttl - explain2}. 
We will therefore derive expressions for these quantities ($R_{m,r}$ and $\varepsilon_{m,r}$) in the following.\\

The radii of curvature are defined via
\begin{equation}
R_{m,r} = z_{m,r} \left(1+\left(\frac{z_0}{z_{m,r}}\right)^2\right)\,,
\label{eq:Rmr}
\end{equation}
where $z_0$ denotes the Rayleigh range. 
The distances from the waists ($z_{m,r}$) can be related to $\delta_{m,r}$ via
\begin{equation}
	z_{m,r} = z+\delta_{m,r} \,,
\end{equation}
where $z$ is the distance from waist in the nominal non-tilted case, defined for the point of incidence at the detector.
We can now substitute $z_{m,r}$ in \cref{eq:Rmr}. Using that $\delta_{m,r}\ll 1$ holds for the typical cases of interest, the radii of curvature can be series expanded to
\begin{equation}
	R_{m,r} \approx R + \delta_{m,r} \left(2-\frac{R}{z}\right) \, .
\label{eq: system of eq radius}
\end{equation}
Here,
\begin{eqnarray}
	R=z \left(1+\left(\frac{z_0}{z}\right)^2\right)
\end{eqnarray}
is the radius of curvature at the point of incidence of a non-tilted beam.\\

In the next step, we define the auxiliary lengths $\varepsilon_{m,r}$. As shown in Fig.~\ref{fig: compute vanishing ttl - explain2}, they describe the distance between the centre of rotation (CoR) and the respective centre of wavefront curvature (CoC$_{m,r}$). 

Any point on the detector can be described by an angle $\psi$ in relation to the initial point of incidence as shown in the right-hand sketch of Fig.~\ref{fig: compute vanishing ttl - explain2}.
There, the point $P$ of interest, the centre of rotation and the centres of curvature span triangles.
By geometric relations (law of cosines) we then know that
\begin{eqnarray}
  R_{m,r}^2 =&\ d_\text{lever}^2 \sec^2(\psi) + \varepsilon_{m,r}^2 \nonumber\\
& - 2d_\text{lever}\sec(\psi)\,\varepsilon_{m,r}\cos(\varphi_{m,r}-\psi) \;.
\label{eq: system of eq trigonometry}
\end{eqnarray}
We have now derived expressions for $R_{m,r}$, $\varepsilon_{m,r}$, $\delta_{m,r}$ (i.e. \eqref{eq: system of eq epsilon},\eqref{eq: system of eq radius},\eqref{eq: system of eq trigonometry}) and can solve these for $\delta_{m,r}$. In order to show the symmetry of the phase difference of the interfering beams and the cancellation of the LPS signal, we derive $\delta_{m,r}$ for points $P_1$ and $P_2$ that are symmetric around the point $x_c$, see Fig.~\ref{fig: compute vanishing ttl}.  In the next step, we then argue that the deviations $\delta_{m,r}$ are equal in the symmetry point $x_c$, such that the difference is zero.

We solve the system of equations \eqref{eq: system of eq epsilon},\eqref{eq: system of eq radius},\eqref{eq: system of eq trigonometry} for $\delta_{m,r}$ studying two points on the detector that possess in a good approximation the same distance to the center of the points of incidence, i.e. $\psi_1=\varphi_m-\psi$ and $\psi_2=\psi$. We get
\begin{eqnarray}
\delta_{1r} &\approx \frac{d_\text{lever}^2}{2R}(\varphi_m-\psi)^2 \,, \\
\delta_{1m} &\approx \frac{d_\text{lever}}{2}(\varphi_m^2-2\varphi_m\psi) + \frac{d_\text{lever}^2}{2R}\psi^2 \,, \\
\delta_{2r} &\approx \frac{d_\text{lever}^2}{2R} \psi^2 \,, \\
\delta_{2m} &\approx \frac{d_\text{lever}(d_\text{lever}-R)}{2R}(\varphi_m^2-2\varphi_m\psi) + \frac{d_\text{lever}^2}{2R}\psi^2 \,,
\end{eqnarray}
and therefore
\begin{eqnarray}
 \delta_{1m}-\delta_{1r} &= -\frac{d_\text{lever}(d_\text{lever}-R)}{2R}(\varphi_m^2-2\varphi_m\psi) \\
&= -(\delta_{2m}-\delta_{2r}) \, .
\label{eq: vanishing ttl symmetric}
\end{eqnarray} 
This shows that the phase differences between measurement and reference beam are identical in the symmetric points $P_1$ and $P_2$ but with inverse signs.

In the centre point $x_c$, which can be described via $\psi = \varphi_m/2$, we find a phase difference of zero in Eq.~\eqref{eq: vanishing ttl symmetric}. This special case is additionally visualised in Fig.~\ref{fig: compute vanishing ttl - explain}.
In this figure, we see that the line connecting the CoR with $x_c$ is (for small angles) the bisector. For symmetry reasons, the wavefronts that are intersecting along this bisecting line must belong to beams having propagated the same distance. It follows $z_m=z+\delta_m=z+\delta_r=z_r$ and hence $\delta_m-\delta_r=0$. \\

Taking together our findings, we have now shown that the amplitude profile of the interference pattern is symmetric to the plane defined by $x=x_c$, where $x_c$ is the centre between the points of incidence along the $x$-axis. Furthermore, the phase differences at points symmetrically arranged to this plane have identical magnitude but inverse signs.
In the LPS signal, the amplitude scales the phase difference. Furthermore, the interferometric LPS signal experimentally originates from the complete detector surface, which mathematically is described by an integral over all detector points (see e.g. \cite{wanner2012methods, Wanner2014}). Therefore, the shown symmetries result in a cancellation of the LPS signal.

A detailed mathematical derivation of the LPS signal for the setup described here and the cancellation in the integral is additionally shown in \cite{NG21}.

\bibliographystyle{unsrt}

\end{document}